\documentclass[5p]{article}

\usepackage{graphicx}
\usepackage{amssymb,amsthm}
\usepackage{color}

\begin{document}

\resizebox{!}{0.4cm}{\bf Symplectic Maps for  Diverted Plasmas}
\\
\\
I. L. Caldas$^1$, B. F. Bartoloni$^2$, D. Ciro$^3$, G. Roberson$^4$, A. B. Schelin$^5$, Tiago Kroetz$^6$, Marisa Roberto$^4$, Ricardo L. Viana$^3$, Kelly C. Iarosz$^1$, Antonio M. Batista$^7$, Philip J. Morrison$^8$
\\
$^1$Instituto de F\'isica, Universidade de S\~ao Paulo, S\~ao Paulo, SP, Brazil.\\
$^2$Departamento de Ensino Geral at Faculdade de Tecnologia de S\~ao Paulo, S\~ao Paulo, Brazil.\\
$^3$Departamento de F\'isica, Universidade Federal do Paran\'a, Curitiba, PR, Brazil.\\
$^4$Aeronautics Institute of Technology at CTA, S\~ao Jos\'e dos Campos, Brazil.\\
$^5$Institute of Physics at University of Bras\'ilia, Bras\'ilia, Brazil.\\
$^6$Department of Mathematics at Federal Technological University of Paran\'a, Pato Branco, Brazil.\\
$^7$Departamento de Matem\'atica e Estat\'istica, Universidade Estadual de Ponta Grossa, Ponta Grossa, PR, Brazil.\\
$^8$Institute for Fusion Studies at The University of Texas at Austin, USA.\\
\noindent Corresponding author: ibere@if.usp.br

\date{\today}

\begin{abstract}
Nowadays, divertors are used in the main tokamaks to control the magnetic field and to improve the plasma confinement. In this article, we present analytical symplectic maps describing Poincar\'e maps of the magnetic field lines in  confined plasmas with a single null poloidal divertor. Initially, we present a divertor map and the tokamap for a  diverted configuration. We also introduce the Ullmann map for a diverted plasma, whose control parameters are determined from tokamak experiments. Finally, an explicit, area-preserving and integrable magnetic field line map for a single-null divertor tokamak is obtained using a trajectory integration method to represent toroidal equilibrium magnetic surfaces. In this method, we also give examples of onset of chaotic field lines at the plasma edge due to resonant perturbations.
\end{abstract}

keywords: magnetic surfaces, sympletic map, divertor



\section{Introduction}
Tokamaks are the most promising devices to confine fusion plasmas
\cite{iter2015, turbulent2012}. The plasma confinement depends on the magnetic field which
determines the particle transport \cite{hazeltine2003}. To the leading order
approximation, the charged particles follow the magnetic field lines
\cite{turbulent2012, hazeltine2003}. Thus, the particle transport can be controlled by properly
modifying the magnetic field as a result of electrical currents in external coils and also by
installing poloidal divertors \cite{iter2015, karger1977}. Such divertors are used to control the
plasma impurity content \cite{wagner1982, kikuchi2012} and have a special magnetic configuration
created by electric currents in external coils, such that the field lines have escape channels,
through which plasma particles can be diverted out of the tokamak wall and redirected to divertor
plates. 

Divertors are essential components in modern tokamaks, such as ITER \cite{iter2015, iter1999}. The
overlap of the magnetic fields created by the divertor with the magnetic field of the plasma creates a
hyperbolic fixed point where the poloidal magnetic field is null. The hyperbolic point is in the
separatrix, the invariant line separating the plasma, with stable and unstable manifolds \cite{evans2002, silva2002}.
Outside the separatrix the magnetic field lines intersect the collector plates \cite{iter2015}.

The tokamak map trajectories can be obtained by directly integrating the field line differential
equations, but the integration requires a time-consuming calculation which may not be appropriate for studying
long-term of the field behavior. Therefore, approximated maps have to
be considered if one wants to have the advantage of much shorter computation times
\cite{hazeltine2003, morrison2000, abdullaev2006}. Analytical tokamak maps can be derived from
physical models and mathematical approximations applied to the field line equations, or even can be ad
hoc maps to obtain a qualitative or quantitative description of the physical situation that they describe
\cite{abdullaev2006, portela2008, caldas2002}. 

Magnetic field lines are, in general, orbits of Hamiltonian systems of one-and-a-half degrees of
freedom with a time-like periodic coordinate. Consequently, the field line configuration can be
represented in Poincar\'e sections at a fixed toroidal angle, equivalent to
two-dimensional area-preserving maps \cite{hazeltine2003, morrison2000, abdullaev2006}. Thus, we can
use such maps to qualitatively represent the magnetic configurations of tokamak plasmas
\cite{barocio2006}. In this work, we present maps proposed to investigate
the fundamental features of the magnetic field line dynamics in tokamaks with divertor. We introduce
new versions of the Toakamap and Ullmann map for tokamaks with divertor and review the Divertor
Map and an integrable map for equilibrium divertor configuration in toroidal geometry.

The paper is organized as follows: in Section \ref{sec2}, we show the tokamak divertors. In
Section \ref{sec3}, we introduce three symplectic maps: the Divertor Map, the Tokamap
for divertor configuration, and the Ullmann map for divertor. We also give examples of these maps to
show their dynamical characteristics. In Section \ref{sec4}, we introduce an integrable map to
simulate toroidal magnetic surfaces modified by a divertor. Section \ref{conclusion} contains the
conclusions.


\section{Divertor}\label{sec2}
In tokamaks, a material limiter separates the plasma column from the wall. However, to improve
the plasma isolation and eliminate impurities, divertors have been used in several modern tokamaks
and will be used in ITER \cite{iter1999}. Divertors consist of conductors arranged externally, that
carry specific electric currents to create $X$ point (or hyperbolic fixed point) where the poloidal
magnetic field is null, due to the overlap of the magnetic fields of the conductors with the magnetic
field of the plasma.

\begin{figure}[!t]
\centering
\includegraphics[width=1.5in]{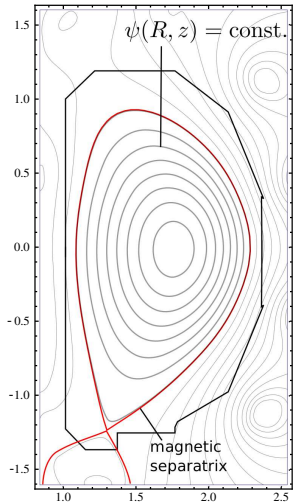}
\caption{(Color online) Schematic view of a tokamak divertor equilibrium configuration with intersection of invariant
  magnetic surfaces on a plane determined by a specific toroidal angle. The red line indicates the magnetic separatrix
  while the contours are magnetic surfaces.}
\label{fig1}
\end{figure}

In Fig. (\ref{fig1}), we present an example of the magnetic surfaces in a tokamak with divertor. In
this figure, the separatrix in red, with one hyperbolic point, separates the internal toroidal
magnetic surfaces with quasi-periodic lines of the external surfaces with open field lines. Moreover,
from the $X$ point arises a separatrix with two manifolds, one stable and the other unstable. 

Several tokamaks with divertors have non-axisymmetric resonant perturbation coils designed specifically to modify the
plasma magnetic field \cite{iter2015, turbulent2012}. One of the actions of the resonant perturbations
created by these coils is to create chaotic magnetic field layers in the peripheral region of the
plasma column \cite{iter1999, abdullaev2006, caldas2002}. 

\begin{figure}[!t]
\centering
\includegraphics[width=2.5in]{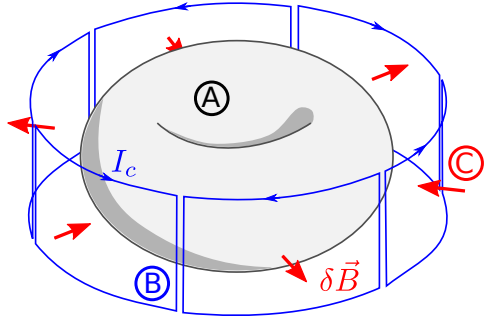}
\caption{(Color online) Schematic view of a tokamak vacuum chamber (A), and the external coils (B),
  responsible for the resonant magnetic field (C). Blue lines indicate the perturbing external currents
  and the red vectors the non-axisymmetric field perturbation.}
\label{fig2}
\end{figure}

\begin{figure}[!t]
\centering
\includegraphics[width=3.0in]{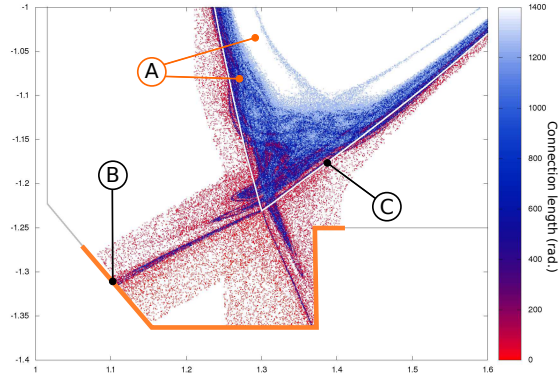}
\caption{(Color online) Detail of the Poincar\'e map of the perturbed field lines in the magnetic saddle region. The
  magnetic perturbation leads to the formation of a peripheral chaotic layer and magnetic islands (A). Chaotic field
  lines now cross the symmetric separatrix (C), and the open field lines intersect the tokamak chamber (B) in asymmetric
  patterns controlled by the invariant manifolds of the saddle.}
\label{fig3}
\end{figure}

In Fig. (\ref{fig2}), we show an example of this kind of coils arranged around a tokamak chamber,
similar to the coils used in DIII-D tokamak \cite{ciro2017}. To show how the tokamak equilibrium is
perturbed by the coils of Fig. (\ref{fig2}), we present in Fig. (\ref{fig3}), the transversal cross
section of the diverted tokamak magnetic field lines for a set of control parameters commonly found
in tokamak discharges \cite{ciro2017, martins2014}. In Fig. (\ref{fig3}) we can see chaotic lines
and magnetic islands around the divertor hyperbolic point, the separatrix of the
unperturbed diverted field, and the divertor plates where the chaotic lines intersect the tokamak
chamber.

The chaotic layer at the plasma edge affects the plasma confinement \cite{vannucci1989} and can be
controlled by the perturbation introduced by the divertor \cite{iter1999}. The chaotic layer in this region is mainly
determined by the manifolds from the hyperbolic point \cite{wingen2009}.

\section{Sympletic Maps} \label{sec3}
Symplectic maps have been commonly used in physics to describe Poincar\'e sections of dynamical systems
\cite{meiss1992, greene1979, chirikov}. In plasma physics, a pioneer symplectic map to describe
particle orbits for stellarators was introduced in \cite{kruskal1952}. After that, symplectic maps
have been used to investigate particle transport in magnetically confined plasmas
\cite{horton1998, castilho1993, castilho1996}. Symplectic maps have also been introduced to
investigate the chaotic field lines in tokamaks. The first one was the Martin-Taylor map introduced
to describe the perturbation created by the ergodic magnetic limiter in tokamaks \cite{martin1984}.

In this section, we present symplectic maps to describe the diverted magnetic fiel lines in tokamaks:
the Divertor Map introduced in \cite{punjabi1992} and new versions of the Tokamap \cite{balescu1998}
and the Ullmann's map \cite{ulmann2000} for divertors. These maps are nonintegrable and describe
plasma equilibrium with chaotic layers, around the hyperbolic point, due to resonant perturbations. 

\subsection{Divertor Map} \label{sec3a}
The first Divertor Map has been presented as the simplest model for the magnetic configuration of a
tokamak equipped with a divertor. The map simulates Poincar\'e sections of field lines
\cite{punjabi1992}.

The Divertor Map introduced in \cite{punjabi1992} is
\begin{eqnarray} \label{divmap}
x_{\rm n+1}&=&x_{\rm n}-ky_{\rm n}(1-y_{\rm n}), \\
y_{\rm n+1}&=&y_{\rm n}+kx_{\rm n+1},\nonumber
\end{eqnarray}
where ($x_{\rm n}$, $y_{\rm n}$) are the rectangular coordinates on the poloidal surface of section and
the control parameter $k$  determines both the safety factor and the strength of toroidal
asymmetries in the magnetic field.  In this map, the equilibrium and perturbation are not separable.

\begin{figure}[!t]
\centering
\includegraphics[width=3.0in]{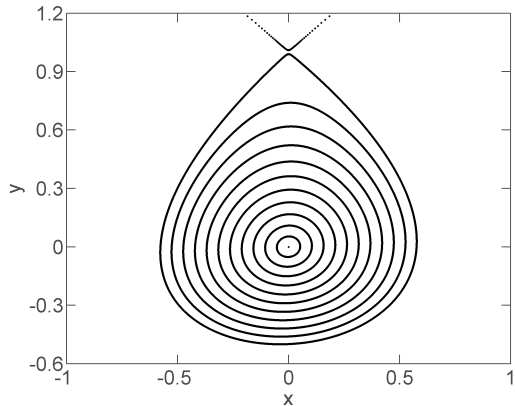}
\caption{Poincar\'e map of the Divertor Map for $k = 0.6$, depicting the stable fixed point at (0,0) and
  some invariant curves. Closed invariant lines are separated from open lines (not shown) by the
  separatrix,  which includes the hyperbolic point at (0,1).}
\label{fig4}
\end{figure}

For small values of the control parameter $k$, the map shows the formation of a thin chaotic layer in
the separatrix region, whose chaotic orbits eventually reach the plates, which are set in the
numerical simulations at $y_{\rm plate}=1$. One example is in Fig. (\ref{fig4}) for $k=0.6$.

\subsection{Tokamap for Diverted Plasmas} \label{sec3b}
The Tokamap has been introduced to describe field lines of a tokamak equilibrium modified by a
resonant perturbation. There are several versions of this map to account for different equilibria and
perturbations. Here, we consider the following version of the map for tokamak plasmas
\cite{balescu1998}:
\begin{eqnarray}
\label{tokamakmap}
\psi_{\rm k}&=&\psi_{\rm k}-\frac{L}{2\pi}\frac{\psi_{\rm k+1}}{1+\psi_{\rm k+1}}\sin(2\pi \theta_{\rm k}),\\
\theta_{\rm k+1}&=&\theta_{\rm k}+\frac{1}{q(\psi_{\rm k+1})}-\frac{L}{2\pi}\frac{1}{(1+\psi_{\rm k+1})^{2}}\cos(2\pi \theta_{\rm k}),\nonumber
\end{eqnarray}
where $L$ is a control parameter that simulates the resonant perturbation amplitude and $q$ is the
safety factor that characterizes the tokamak equilibrium.

Here, we consider a safety factor profile $q$, as considered in \cite{kroetz2010}, with a
singularity at $r=r^{*}$. This profile is polynomial for $0 \leq r < r_{\rm 95}$ and logarithmic for
$r_{\rm 95} < r \leq r^{\rm *}$, where $r_{\rm 95}=0.95r^{*}$:
\begin{eqnarray}
q_{\rm 0}(r)&=&q_{\rm a}+c_{\rm 1}r+c_{\rm 2}+r^{\rm 2}, \qquad {\rm if} \quad r\leq r_{\rm 95}, \label{r95}\\
q_{\rm 0}(y)&=& \alpha \ln(y^*-y)+ \beta, \qquad {\rm if} \quad r_{\rm 95} \leq r. \label{q95}
\end{eqnarray}

\begin{figure}[!t]
\centering
\includegraphics[width=3.0in]{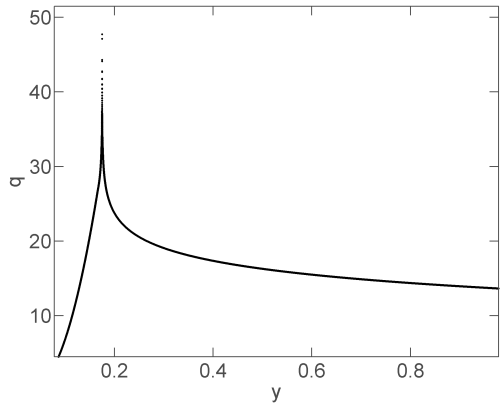}
\caption{Safety factor profile considered for the  the Tokamap to obtain Fig. (\ref{fig6}). The coordinate $y$ corresponds to the radial coordinate used in large aspect ratio tokamaks.}
\label{fig5}
\end{figure}

\begin{figure}[!t]
\centering
\includegraphics[width=3.0in]{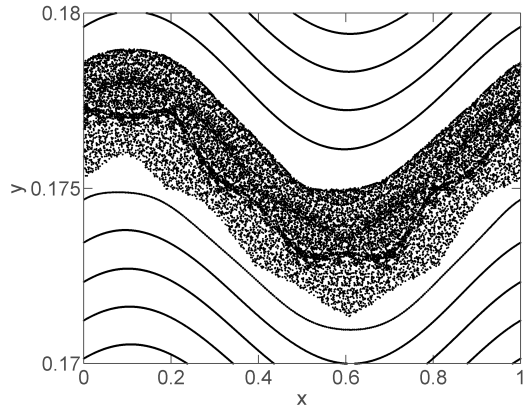}
\caption{Invariant magnetic surfaces and chaotic field lines obtained from the Tokamap for
  L = 0.1.}
\label{fig6}
\end{figure}

We choose the q profile parameters such that: $q_{\rm 0}$=11.1, $q_{\rm 95}$=3.3, $r^{*}$=0.18,
$r_{\rm 95}$=0.171, and $r_{\rm 95}, s_{\rm 95}$=110.8 is related to magnetic shear for the reference surface
$r_{\rm 95}$. The considered profile is shown in Fig. (\ref{fig5}). In Fig. (\ref{fig6}), we show 
invariant and chaotic lines, around the hyperbolic point, for the control parameter L=e$^{-1}$.
The observed chaotic layer appers in the resonant region.

\subsection{Ullmann Map for Diverted Plasma} \label{sec3c}
A special set of coils, known as the ergodic limiter, have been proposed to create a chaotic layer at
the tokamak plasma edge in order to separate the plasma from the wall \cite{karger1977}. Since then
different kinds of limiters were installed in tokamaks to control the plasma
confinement \cite{portela2008, caldas2002, ghendrih1996}.

In \cite{ulmann2000} a symplectic map was proposed to describe tokamak field lines perturbed by an
ergodic limiter. The map is valid for large aspect ratio tokamaks with toroidal correction. In this
approximation, the toroidal coordinate is related to $z$. In the model, the topology of the magnetic
field lines is described by a Poincar\'e map in the section $z={\rm constant}$, with variables
$r_{\rm n}, \Theta_{\rm n}$ denoting the coordinates of the $n$th intersection of the field line on the
considered section \cite{portela2008, ulmann2000}. 

The analytical expressions for the Poincar\'e map is obtained, for the equilibrium with toroidal
correction, by the generating function:
\begin{eqnarray}
\label{poincarecylindrical}
G_{\rm TO}(r_{\rm n}, \theta_{\rm n})=G_{\rm cil}(r_{\rm n},\theta_{\rm n})+\sum_{\rm l=1}^{\infty}a_{\rm l}\left (\frac{r_{\rm n}}{R_{\rm 0}} \right)^{\rm l} \cos^{\rm l} \theta_{\rm n}
\end{eqnarray}
and the following relations:
\begin{eqnarray}
  r_{\rm n}&=&\frac{\partial G_{\rm TO}(r_{\rm n}, \theta_{\rm n})}{\partial \theta_{\rm n}}, \label{rn}\\
  \theta_{\rm n}&=&\frac{\partial G_{\rm TO}(r_{\rm n}, \theta_{\rm n})}{\partial r_{\rm n}}. \label{thetan}
\end{eqnarray}

From equations (\ref{rn}) and (\ref{thetan}) we obtain the map expressions
\begin{eqnarray}
  r_{\rm n+1}&=&\frac{r_{\rm n}}{1-a_{\rm 1}\sin \theta_{n}}, \label{rn+1}\\
  \theta_{\rm n+1}&=&\theta_{\rm n}+\frac{2\pi}{q(r_{\rm n+1})}+a_{\rm 1}\cos \theta_{\rm n}, \label{thetan+1}
\end{eqnarray}
where $a_{\rm 1}$ is a correction due to the toroidal effect and $q(r)$ is the safety factor profile.
The toroidal correction introduces a poloidal angle $\theta$ dependence on the map. Such correction,
considered in the model, takes into account the outward magnetic surface displacement, characteristic
of tokamak equilibria in toroidal geometry. The constant $a_{\rm 1}=-0.04$ was fit to reproduce
the observed tokamak magnetic surface displacements \cite{ulmann2000}. Since the map is derived from
a generating function, interpreted as a canonical transformation between the previous and the next
coordinates, the Jacobian for this map is unitary and, consequently, the map is symplectic
\cite{portela2008, ulmann2000}.

We consider the same safety factor profile $q_{\rm 0}$ considered for the Tokamap, with a singularity
in the same position and the same parameters $q_{\rm 0}=11.1$, $q_{\rm 95}=3.3$, $r^{\rm *}=0.18$,
$r_{\rm 95}=0.171$ and the magnetic shear, for the reference surface $r_{\rm 95}, s_{\rm 95}=110.8$.

Neglecting the toroidal correction, i.e., for $a_{\rm 1}=0$, the map is integrable and the rotation
number profile is the inverse of the safety factor $q_{\rm 0}$ appearing in equations (\ref{r95}) and
(\ref{q95}) with a divergence at $r=r^{\rm *}$. However, including the predicted toroidal correction and
considering $a_{\rm 1}=-0.04$, mentioned before, the rotation number has to be calculated numerically
by the expression
\begin{eqnarray}
\label{rotationnumber}
q=\frac{1}{l} \rightarrow q \equiv \lim_{k \rightarrow \infty}\frac{2 \pi k}{\sum_{j=0}^{k}(\theta_{\rm j+1}-\theta_{\rm j})}.
\end{eqnarray}

\begin{figure}[!t]
\centering
\includegraphics[width=3.0in]{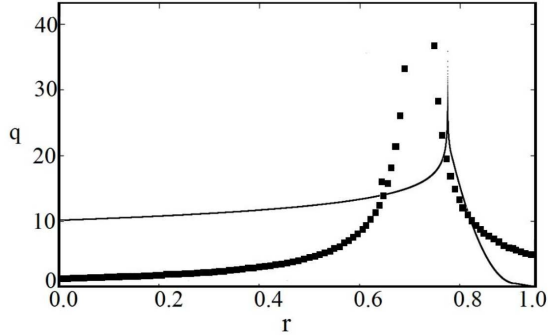}
\caption{Analytical safety factor profile, obtained from equations (\ref{r95}) and (\ref{q95}),
  with a divergence at $r=r^{\rm *}$, and the safety factor profile calculated numerically,
  for $\theta=y=0.05$. The coordinate $r$ is normalized to $a$ (plasma radius).}
\label{fig7}
\end{figure}

In Fig. (\ref{fig7}), we have the analytical safety factor profile  obtained from this definition
and,  for initial conditions with a fixed $\theta$, the modified safety factor calculated numerically
for 100 values of $r$ between 0 and 1. We see in Fig. (\ref{fig7}) the difference between the original
profile inserted in equation (\ref{thetan+1}) and the one calculated, considering the toroidal
correction, by applying equation (\ref{rotationnumber}). As shown in \cite{portela2008} and
\cite{ulmann2000}, we can add another symplectic map as an external perturbation to obtain a
chaotic layer on the resonant region.

\section{Integrable Map for toroidal Magnetic Surfaces} \label{sec4}
In this section, we introduce a procedure to obtain an integrable map simulating a plasma equilibrium for
a diverted tokamak \cite{kroetz2010, kroetz2012, roberson2017}. 

To obtain the desired map it is necessary, initially, to find a potential $V(x)$ that produces a
topology with an X-point, with a Hamiltonian $\psi$ given by
\begin{eqnarray}
\label{hamiltonian}
\psi=\frac{y^{\rm 2}}{2}+V(x).
\end{eqnarray}

\begin{figure}[!t]
\centering
\includegraphics[width=5in]{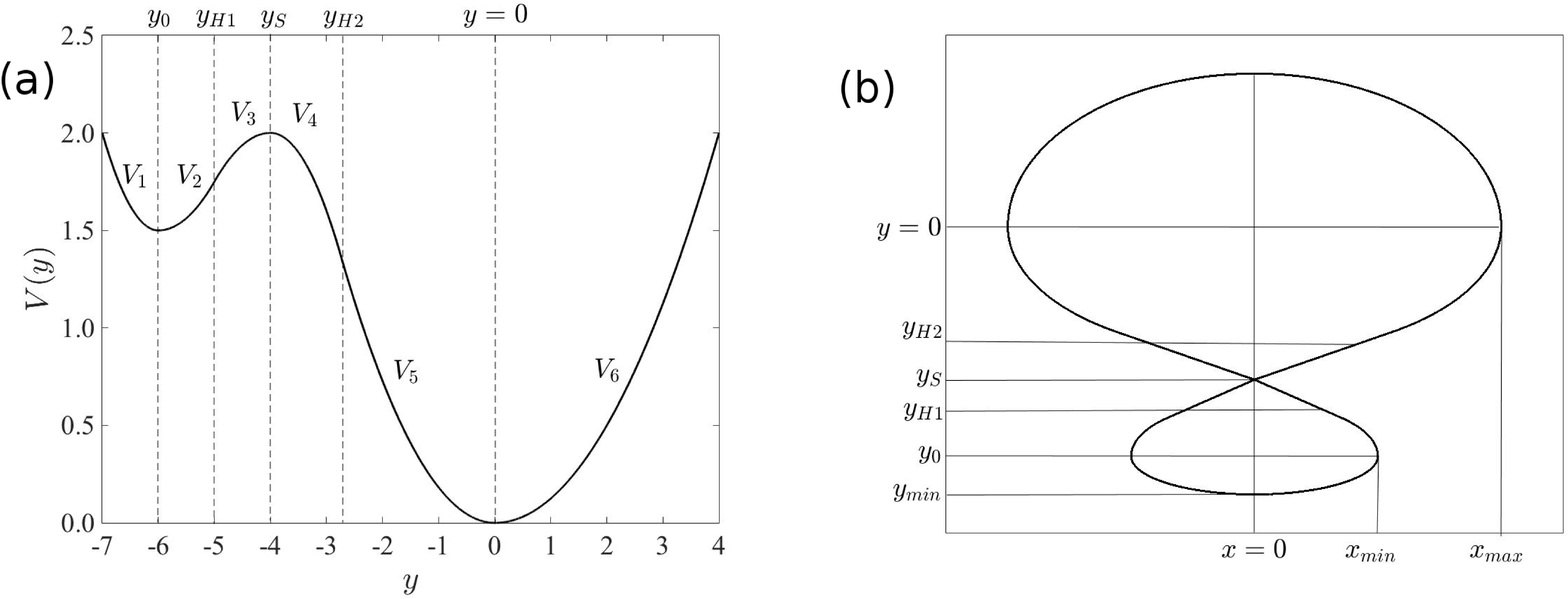}
\caption{(a) Potential used to obtain the magnetic surfaces. (b) Schematic view of a separatrix in
  rectangular coordinates indicating the meaning of each geometric parameter related to $V(y)$ for the
  normalized geometric parameters $x_{\rm max}=-2$, $x_{\rm min}=-1$, $y_{min}=−7$, $y_{\rm max}=4$, $y_{\rm H1}=−5$,
  $y_{\rm H2}=−2.75$, $y_{\rm S}=−4$, $y_{\rm 0}=−6$.}
\label{fig8}
\end{figure}

In the plasma map this Hamiltonian will be the flux function. We choose a double-well shaped potential
to create curves in phase space that exhibit two closed regions delimited by a separatrix between
them with an $X$ point. The expression for $V(x,y)$ with the desired properties will be written as a set
of six parabolas, indicated in Fig. (\ref{fig8}a), joined smoothly the connection points.
Figure (\ref{fig8}a) shows the chosen potential profile used in this article \cite{kroetz2012}. The
position $y=0$ corresponds to the plasma center. In Fig. (\ref{fig8}b), we present the separatrix
for the orbits obtained for the chosen potential of Fig. (\ref{fig8}a). This separatrix 
determines, in the map, the last closed magnetic surfaces inside the plasma.

We choose an analytical expression for the potential represented in Fig. (\ref{fig8}a). For this
potential, the trajectory corresponding to the separatrix is shown in Fig. (\ref{fig8}b), for ITER
parameters.

The next step is to solve Hamilton's equations to get $x$ and $y$ in terms of their initial
conditions $(x_{\rm 0},y_{\rm 0})$ and time $t$,
\begin{eqnarray}
  \label{solveH}
  \frac{dx}{dt}&=&-\frac{\partial \psi}{\partial y},\\
  \frac{dy}{dt}&=&\frac{\partial \psi}{\partial x}.
\end{eqnarray}

The continuous equations are transformed into a discrete map, where the continuous time parameter
$t$ is turned into a discrete time step $\Delta$:
\begin{eqnarray}
  \label{s}
  x(x_{0},y_{0},t)=x_{\rm n+1}(x_{\rm n},y_{\rm n}, \Delta),\\
  y(x_{0},y_{0},t)=y_{\rm n+1}(x_{\rm n},y_{\rm n}, \Delta).
\end{eqnarray}

To obtain the magnetic surfaces we choose $\Delta$ given  by the inverse of the safety factor 
\begin{eqnarray}
  \label{delta}
 \Delta=\frac{T(\psi)}{q(\psi)}.
\end{eqnarray}
where $T(\psi)$ is the rotation period of the invariant curves  associated with the continuous system and
$q(\psi)$ is the safety factor of the magnetic surface we intend to represent by the invariant curve \cite{kroetz2012, roberson2017}.

We choose a monotonic safety factor profile similar to the one used before in Section \ref{sec3} (B)
and (C) expressed in terms of the function $\psi$ \cite{kroetz2012}:

$q(\psi)=\begin{cases}
q_{\rm 0}+c_{\rm 1}\psi+c_{\rm 2}\psi^{2}, \qquad \psi \leq \psi_{\rm 95},\\
\alpha \ln (\psi_{\rm S}-\psi)+ \beta, \qquad \psi > \psi_{\rm 95}.
\end{cases}$

In the numerical examples we choose the safety factor parameters such that the magnetic shear at the
reference surface, defined as
\begin{eqnarray}
\label{s95}
\hat{s}_{\rm 95}=\frac{r_{\rm 95}}{q_{\rm 95}}\frac{dq}{dr}\Big|_{\rm r_{95}},
\end{eqnarray}
are $\hat{s}_{\rm 95}=110.8$ and $q_{\rm 95}=3.3$. 

For each line, the value of $\psi$ is given by $\psi=\psi(x_0,y_0)$. At each point ($x_0,y_0$), $\psi$
determines the $\Delta$ value:
\begin{eqnarray}
  \label{deltapsi}
\Delta=\Delta(\psi).
\end{eqnarray}

\begin{figure}[!t]
\centering
\includegraphics[width=3.0in]{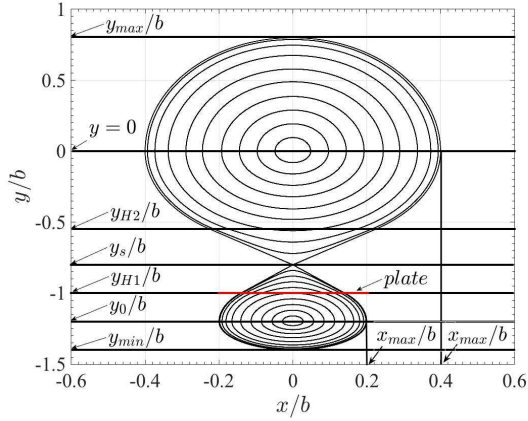}
\caption{(Color online) Invariant magnetic surfaces obtained from the integrable map  for the chosen parameters
  indicated in the text.}
\label{fig9}
\end{figure}

The map gives the Poincar\'e map on the surface $\varphi=0$
\begin{eqnarray}
  \label{deltapsi1}
M_{\Delta}(x_{\rm n},y_{\rm n})=(x_{\rm n+1},y_{\rm n+1}).
\end{eqnarray}
The magnetic surfaces are shown in Fig. (\ref{fig9}).

To perturb the divertor integrable map, we apply the symplectic Martin-Taylor map \cite{martin1984},
that simulates the effect of an ergodic limiter in large aspect-ratio tokamaks, which introduces
external symmetry-breaking resonances, so as to generate a chaotic region near the separatrix
passing through the X-point.

For each toroidal turn, the Martin Taylor map is applied at the $\varphi=0$ surface as a kick
perturbation:
\begin{eqnarray}
  \label{perturbation}
  (x^{*},y^{*})=M_{\rm \Delta_{\rm n}}(x_{\rm n},y_{\rm n}),\\
  (x_{\rm n+1},y_{\rm n+1})=P(x^{*},y^{*}).  
\end{eqnarray}

\begin{figure}[!t]
\centering
\includegraphics[width=5in]{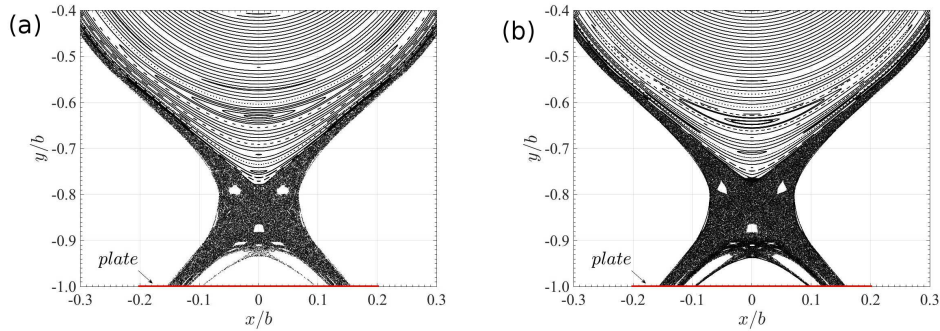}
\caption{(Color online) Phase portrait for the total field line map. (a) $s=1.9$ and $p=3$, (b) $s=2.5$ and $p=3$.}
\label{fig10}
\end{figure}

Thus, the map used to describe the perturbation of an external resonant helical perturbation due
to a magnetic limiter is given by \cite{martin1984}
\begin{eqnarray}
  \label{helical}
  x_{\rm n+1}&=&x_{\rm n}-ce^{-\frac{my_{\rm n}}{r_{\rm c}}}\cos \left(\frac{mx_{\rm n}}{r_{\rm c}}\right),\\
  y_{\rm n+1}&=&y_{\rm n}+\frac{r_{\rm c}}{m} \log \left\{ \cos \left[\frac{mx_{\rm n}}{r_{\rm c}}-ce^{-\frac{mx_{\rm n}}{r_{\rm c}}}\cos \left(\frac{mx_{\rm n}}{r_{\rm c}}\right)\right]\right\}\nonumber \\
  &-&\frac{r_{\rm c}}{m} \log \left\{ \cos \left(\frac{mx_{\rm n}}{r_{\rm c}} \right) \right\},
\end{eqnarray}
where the parameter $c$ quantifies the perturbation strength, proportional to the current in the
limiter coils, $s$ is the magnetic shear at the plasma edge, and $r_{\rm c}$ is the palsma radius.
The composed field line map is used to obtain the perturbed field line configurations, shown
in Fig. (\ref{fig10}), with different magnetic shear profiles at the plasma edge for the control
parameters $s=1.9$ and $s=2.5$ for $c=3$. The current in the ergodic limiter is the
same in Figs. (\ref{fig10}a) and (\ref{fig10}b) .

The introduced non-axisymmetric stationary magnetic perturbatiom leads to the formation of homoclinic
tangles near the divertor magnetic saddle \cite{wingen2009}. These tangles intersect the divertor
plates in static helical structures.

\section{Conclusions} \label{conclusion}
To describe Poincar\'e sections of diverted magnetic field lines, we presented two-dimensional
symplectic maps, in the limit of large aspect ratio simulating the alterations the magnetic
topology caused by the divertor.

These maps can be used to investigate the main characteristics of the chaotic layer around the hyperbolic
point introduced by divertors, and how these characteristics change with the equilibrium and perturbation
control parameters. The Tokamap and Ullmann map were presented in new versions for tokamaks with divertor.
We also presented a map describing magnetic surfaces in toroidal geometry.

All the  maps introduced in this article are useful for studying different aspects of the field line dynamics
and transport in tokamaks with divertor. Extensions of the presented maps could be derived to include
additional effects not considered in this article, such as the particle's finite Larmor radius
\cite{martinell2013, fonseca2014} and the screening caused by the plasma response to resonant magnetic
perturbations \cite{wingen2015, fraile2017}.


\section*{Acknowledgments}
This work was partially supported by the National Council for Scientific and Technological Development (CNPq)-Brazil, grants 870198/1997-1 and 830577/ 1999-8 and the S\~ao Paulo Research Foundation (FAPESP) grants 2012/18073-1, 2015/07311-7 and 2011/19296-1. PJM was supported by the U.S. Department of Energy Contract No.DE-FG05-80ET-53088.


\begin{frame}

\end{frame}
\end{document}